\documentclass[aps,prd,onecolumn,groupedaddress,showpacs,nofootinbib,amssymb]{revtex4}
\usepackage[dvips]{graphicx,color}
\usepackage{amssymb}
\usepackage{amsmath}
\usepackage{graphicx}
\usepackage{amsfonts}

\newcommand{\be}{\begin{equation}}
\newcommand{\ee}{\end{equation}}
\newcommand{\bea}{\begin{eqnarray}}
\newcommand{\eea}{\end{eqnarray}}
\newcommand{\beaa}{\begin{eqnarray*}}
\newcommand{\eeaa}{\end{eqnarray*}}

\begin{document}

\title{Reconstructing the Universe Evolution from Loop Quantum Cosmology Scalar Fields}
\author{V.~K.~Oikonomou$^{1,2}$\,\thanks{v.k.oikonomou1979@gmail.com}}
\affiliation{$^{1)}$ Tomsk State Pedagogical University, 634061 Tomsk, Russia\\
$^{2)}$ Laboratory for Theoretical Cosmology, Tomsk State University of Control Systems
and Radioelectronics (TUSUR), 634050 Tomsk, Russia\\
}

\begin{abstract}
We extend the scalar-tensor reconstruction techniques for classical cosmology frameworks, in the context of loop quantum cosmology. After presenting in some detail how the equations are generalized in the loop quantum cosmology case, we discuss which new features and limitations does the quantum framework brings along, and we use various illustrative examples in order to demonstrate how the method works. As we show the energy density has two different classes of solutions, and one of these yields the correct classical limit while the second captures the quantum phenomena. We study in detail the scalar tensor reconstruction method for both these solutions. Also we discuss some scenarios for which the Hubble rate becomes unbounded at finite time, which corresponds for example in a case that a Big Rip occurs. As we show this issue is non-trivial and we discuss how this case should be treated in a consistent way. Finally, we investigate how the classical stability conditions for the scalar-tensor solutions are generalized in the loop quantum framework. 
\end{abstract}


\maketitle

\section{Introduction}

The strikingly unexpected observation of the late-time acceleration of the Universe in the late 90's \cite{riess} has set the stage for the construction of alternative theories of gravity to model the Universe. Up to date, it is believed that the Universe experienced two acceleration eras, the early-time acceleration and the late-time acceleration era. One characteristic feature of the early-time acceleration era is the production of a slightly red tilted scale invariant power spectrum of primordial curvature perturbations, which has recently been verified by the Planck data \cite{planck}. One of the successful theories that produce a nearly scale invariant spectrum is the inflationary scenario \cite{inflation1,inflation2,inflation2a,inflation2b,inflation2c,inflation3,inflation4,inflation5,inflation6}. However, an alternative scenario to the standard inflationary paradigm is the big bounce evolution \cite{brande1,bounce1,bounce1a,bounce1b,bounce2,bounce3,bounce4,bounce5,matterbounce1,matterbounce2,matterbounce3,matterbounce4,matterbounce5,matterbounce6,matterbounce7}, in which the initial singularity is avoided and also it is possible to produce a scale-invariant power spectrum. Particularly, it was known for quite some time that an inflationary de Sitter evolution and a contracting cosmological phase with scale factor $a(t)\sim (-t)^{2/3}$, are related by a duality \cite{duality}, and both produce scale invariant spectrum. A well known cosmological bounce which realizes a contracting phase which produces an exactly scale invariant spectrum is the matter bounce scenario \cite{matterbounce1,matterbounce2,matterbounce3,matterbounce4,matterbounce5,matterbounce6,matterbounce7}, which naturally arises in the context of Loop Quantum Cosmology (LQC) \cite{LQC1,LQC2,LQC3,LQC4,LQC5,LQC6,LQC19}, if the matter content consists of a pressureless perfect fluid.

Scalar fields are frequently used in order to describe inflationary theories \cite{inflation1,inflation2,inflation2a,inflation2b,inflation2c,inflation3,inflation4,inflation5}, and in order to describe the late-time acceleration era \cite{scalarrecon0,scalarrecon1,scalarrecon2,scalarrecon3,scalarrecon4,scalarrecon5,scalarrecon6,scalarrecon7,scalarrecon8,scalarrecon9}. Particularly, several reconstruction techniques use canonical or non-canonical scalar fields \cite{scalarrecon0,scalarrecon1,scalarrecon2,scalarrecon3,scalarrecon4,scalarrecon5,scalarrecon6,scalarrecon7,scalarrecon8,scalarrecon9} in order to generate a quintessential or even a phantom late-time evolution. The purpose of this paper is to generalize the reconstruction methods of Refs. \cite{scalarrecon0,scalarrecon1,scalarrecon2,scalarrecon3} in the context of LQC. We aim to present the general method of realizing a given cosmological evolution in terms of it's Hubble rate and scale factor, and we investigate how the classical results obtained in Refs. \cite{scalarrecon0,scalarrecon1,scalarrecon2,scalarrecon3} are generalized in the case of LQC. As we will see, the LQC resulting equations are identical to the classical equations, when the classical limit is taken. Both the non-canonical and canonical scalar fields cases shall be discussed, and also we use several illustrative examples to show how the method works. We also discuss the limitations of the method and also we highlight the difference with the classical case. In addition we discuss how the method works in the case that the quantum era is considered, since there are two branches of solutions for the energy density. Also the case that the Hubble rate is unbounded is discussed in brief, in the case of a Big Rip where the unboundedness occurs at a finite time. In the context of LQC, this requires special attention, as was demonstrated in \cite{rev1}, since the Friedmann equations have a different form, however the Big Rip singularity should be avoided as it happens in all LQC frameworks, see for example \cite{nobigrip1,nobigrip1a,nobigrip1b,nobigrip1c,nobigrip2,nobigrip3,nobigrip4,haronobigrip}. We point out the problem of this issue and we discuss how the problem is actually solved by using the arguments of Ref. \cite{rev1}.

This paper is organized as follows: In section II, we discuss the general reconstruction technique for non-canonical LQC scalar fields. In section III we address the stability of the general solution we obtained in section II, and we compare the results with the classical case. In section IV we investigate how phantom and oscillating cosmologies can be generated by LQC non-canonical scalar fields, and we discuss the case of unbounded Hubble rates at finite time. The canonical scalar field case is discussed in section V and the concluding remarks along with a discussion of the results follow in the end of the paper.

\section{The LQC Reconstruction Method with non-Canonical Scalar Field}

According to the theoretical framework of LQC \cite{LQC1,LQC2,LQC3,LQC4,LQC5,LQC6,LQC19}, the effective Hamiltonian that describes the quantum aspects of the Universe is
\begin{equation}\label{effhamilt}
\mathcal{H}_{LQC}=-3V\frac{\sin^2(\lambda \beta)}{\gamma^2\lambda^2}+V\rho\, ,
\end{equation}
where $\gamma$ is the Barbero-Immirzi parameter, $\lambda$ is a parameter with dimensions of length, which is related to $\gamma$, $V$ represents the volume $V=a(t)^3$, with $a(t)$ denoting the scale factor of the Universe, and finally $\rho$  represents the energy density of the matter content of the Universe. The Hamiltonian constraint $\mathcal{H}_{LQC}=0$, yields,
\begin{equation}\label{hamiltonianconstr}
\frac{\sin^2(\lambda \beta)}{\gamma^2\lambda^2}=\frac{\rho}{3}\, ,
\end{equation}
and in addition by using the following anti-commutation identity,
\begin{equation}\label{anticom}
\dot{V}=\{V,\mathcal{H}_{LQC}\}=-\frac{\gamma}{2}\frac{\partial \mathcal{H}_{LQC}}{\partial \beta}, 
\end{equation}
we get the holonomy quantum corrected Friedmann equation \cite{LQC1,LQC2,LQC3,LQC4,LQC5,LQC6,LQC19},
\begin{equation}\label{holcor1}
H^2=\frac{\kappa^2\rho}{3}\left (1-\frac{\rho}{\rho_c}\right )\, ,
\end{equation}  
The energy density satisfies as usual the following continuity equation,
\begin{equation}\label{cont}
\dot{\rho}(t)=-3H\Big{(}\rho(t)+P(t) \Big{)}\, ,
\end{equation}
where $P(t)$ is the total effective pressure of the matter fluid with energy density $\rho$. Using the continuity equation (\ref{cont}), the holonomy corrected Friedman equation (\ref{holcor1}), by differentiating the latter, and finally substituting $\dot{\rho}$ from the continuity equation (\ref{cont}), we obtain the following differential equation,
\begin{equation}\label{eqnm}
\dot{H}=-\frac{\kappa^2}{2}(\rho+P)(1-2\frac{\rho}{\rho_c})\, .
\end{equation}
When using the LQC theoretical framework, one of the fundamental guidelines is that the resulting picture in the context of LQC, must yield the classical picture, in the case that the LQC parameter $\rho_c\to \infty$, since the parameter $\rho_c$ captures the quantum effects. Notice that in the limit $\rho_c\to \infty$, the two equations (\ref{holcor1}) and (\ref{eqnm}) become,
\begin{equation}\label{classicaleqns}
\dot{H}=-\frac{\kappa^2}{2}(\rho+P),\,\,\,H^2=\frac{\kappa^2\rho}{3}\, ,
\end{equation}
which are the classical Friedmann equations. Having the LQC Friedmann equations at hand, we now extend the formalism of Refs. \cite{scalarrecon0,scalarrecon1,scalarrecon2} in the case of Loop Quantum Cosmology non-canonical scalar fields. From the LQC Friedmann equation (\ref{holcor1}), the solution with respect to $\rho$ is,
\begin{equation}\label{sol1}
\rho=\frac{\kappa ^2 \rho_c\pm \sqrt{-12 H^2 \kappa ^2 \rho_c+\kappa ^4 \rho_c^2}}{2 \kappa ^2}\, ,
\end{equation}
and we need to be cautious at this point for two reasons: first the limit $\rho_c\to \infty$ must yield the classical solution $H^2=\frac{\kappa^2\rho}{3}$, and secondly we need to choose the correct sign in Eq. (\ref{sol1}). These two problems are related as we now evince. Now we demonstrate which solution we need to keep and we will write the appropriate expression for $\rho$ in order for the classical solution to be obtained. The total pressure density from Eq. (\ref{eqnm}) reads,
\begin{equation}\label{pr}
P(t)=-\rho-\frac{2\dot{H}}{\kappa^2(1-\frac{2\rho}{\rho_c})}\, ,
\end{equation}
therefore, the corresponding Equation of State (EoS) parameter $w_{eff}$ reads,
\begin{equation}\label{eosdef}
w_{eff}=-1-\frac{\dot{H}}{3H^2}\pm \frac{\rho_c \dot{H}}{3 H^2 \sqrt{\rho_c \left(\rho_c-12 H^2\right)}}\, ,
\end{equation} 
where the plus sign corresponds to the plus sign in Eq. (\ref{sol1}) and the minus in (\ref{eosdef}) to the minus in (\ref{sol1}). Obviously, when $\rho_c\to \infty$, the EoS parameter must be equal to the classical result \cite{scalarrecon0,scalarrecon1,scalarrecon2},
\begin{equation}\label{claseos}
w_{eff}=-1-\frac{2\dot{H}}{3H^2}\, ,
\end{equation}
and the only way for this to be true is if we keep only the minus sign in Eq. (\ref{sol1}), so in effect the final expression for the energy density reads,
\begin{equation}\label{rhodefcorrect}
\rho=\frac{\kappa ^2 \rho_c- \sqrt{-12 H^2 \kappa ^2 \rho_c+\kappa ^4 \rho_c^2}}{2 \kappa ^2}\, ,
\end{equation}
and therefore the resulting correct expression for the LQC EoS is the following,
\begin{equation}\label{eosdefcorrect}
w_{eff}=-1-\frac{\dot{H}}{3H^2}- \frac{\rho_c \dot{H}}{3 H^2 \sqrt{\rho_c \left(\rho_c-12 H^2\right)}}\, .
\end{equation}
A second problem arises if we take the limit $\rho_c\to \infty$ in Eq. (\ref{rhodefcorrect}), in which case we get $\rho=0$, which is not the classical result. The source of the problem is that we cannot take the limit $\rho_c\to 0$ directly from Eq. (\ref{rhodefcorrect}), so we write the definition of $\rho$ in an equivalent form using Eqs. (\ref{holcor1}) and (\ref{rhodefcorrect}),
\begin{equation}\label{correctdef}
\rho=\frac{3H^2}{\kappa^2\left( 1-\frac{\kappa ^2 \rho_c- \sqrt{-12 H^2 \kappa ^2 \rho_c+\kappa ^4 \rho_c^2}}{2 \kappa ^2\rho_c}\right)}
\end{equation}
which in the limit $\rho_c\to \infty$, gives the correct result $H^2=\frac{\kappa^2\rho}{3}$. Basically it is a trivial algebraic trick which we obtain by using Eqs. (\ref{holcor1}) and (\ref{rhodefcorrect}). By substituting (\ref{rhodefcorrect}) in Eq. (\ref{correctdef}) we can see that it holds true. So from now on the total energy density is given by Eq. (\ref{correctdef}) while the total pressure from (\ref{pr}) and the EoS parameter $w_{eff}$ from (\ref{eosdef}).

However, we need to stress that the root given in Eq. (\ref{rhodefcorrect}) yields the correct classical limit, but the following root,
\begin{equation}\label{rhodefcorrectrev}
\rho=\frac{\kappa ^2 \rho_c+ \sqrt{-12 H^2 \kappa ^2 \rho_c+\kappa ^4 \rho_c^2}}{2 \kappa ^2}\, ,
\end{equation}
yields the physical description of the theory when the quantum effects are strong. Practically, the root (\ref{rhodefcorrectrev}) describes the bouncing era, which resides to the quantum regime of the theory. Therefore, in the following analysis we need to find the behavior corresponding to the quantum era root of Eq. (\ref{rhodefcorrectrev}). More importantly, both the roots (\ref{rhodefcorrect}) and (\ref{rhodefcorrectrev}) are derived when the Hubble rate is considered to be bounded, and in the contrary case the result is changed, since the Friedmann equation is not given by Eq. (\ref{holcor1}) anymore. A recent insightful study of these issues was performed in Ref. \cite{rev1}, where the authors performed a very general analysis using the Raychaudhuri equation, and their analysis yielded quite novel results. We shall discuss the issue of an unbounded Hamiltonian in a later section.

From the expression of Eq. (\ref{eosdefcorrect}) for the EoS, it is obvious that the EoS describes a phantom evolution when $\dot{H}>0$ and on the contrary, when $\dot{H}<0$, the evolution is non-phantom. In general, we can say that the effect of LQC in the EoS is the appearance of the third term in (\ref{eosdefcorrect}), which in the case that
\begin{equation}\label{den1}
\frac{\rho_c}{ \sqrt{\rho_c \left(\rho_c-12 H^2\right)}}<1\, ,
\end{equation}
 then the evolution is less phantom or more quintessential in comparison to the classical EoS given in Eq. (\ref{claseos}). On the other hand, if  
\begin{equation}\label{den1}
\frac{\rho_c}{ \sqrt{\rho_c \left(\rho_c-12 H^2\right)}}>1\, ,
\end{equation}
then, the evolution becomes more phantom or less quintessential, compared to the classical EoS. In the following we shall make use of a flat Friedmann-Robertson-Walker background, with line element,
\be
\label{metricfrw} ds^2 = - dt^2 + a(t)^2 \sum_{i=1,2,3}
\left(dx^i\right)^2\, .
\ee
Having specified the geometric background, consider the case that the energy density $\rho$ consists of a non-canonical scalar field, with pressure $P_{\phi}$ and energy density $\rho_{\phi}$,
\begin{equation}\label{presenrg}
\rho_{\phi}=\frac{1}{2}\omega(\phi)\dot{\phi}^2+V(\phi),\,\,\,P_{\phi}=\frac{1}{2}\omega(\phi)\dot{\phi}^2-V(\phi)\, ,
\end{equation}
and also of an ordinary matter fluid with equation of state $P_m=w_m\rho_m$, so that,
\begin{equation}\label{completeenergypressure}
P=P_m+P_{\phi},\,\,\,\rho=\rho_m+\rho_{\phi}\, ,
\end{equation}
and it is assumed that the two perfect fluids do not interact. Then, the kinetic term for the LQC non-canonical scalar reads,
 \begin{equation}\label{omega}
 \omega(\phi)\dot{\phi}^2=-2\frac{\dot{H}}{\kappa^2(1-2\frac{\rho}{\rho_c})}-(\rho_m+P_m),
\end{equation}
and the potential is,
\begin{equation}\label{potential}
V(\phi)=\rho+\frac{\dot{H}}{\kappa^2(1-2\frac{\rho}{\rho_c})}-\frac{\rho_m}{2}+\frac{P_m}{2}\, ,
\end{equation}
and in the limit $\rho_c\to \infty$ we obtain the correct classical expressions. By using Eq. (\ref{correctdef}), the potential reads,
\begin{equation}\label{potential1}
V(\phi)= \frac{3H^2}{\kappa^2-\frac{\kappa ^2 \rho_c- \sqrt{-12 H^2 \kappa ^2 \rho_c+\kappa ^4 \rho_c^2}}{2 \rho_c}}+\frac{\dot{H}}{\kappa^2(1-2\frac{\rho}{\rho_c})}-\frac{\rho_m}{2}+\frac{P_m}{2}\, ,
\end{equation}
We seek solutions of the form \cite{scalarrecon0,scalarrecon1,scalarrecon2},
\begin{equation}\label{solutionsoftheform}
\phi=t,\,\,\,H(t)=f(t)\, ,
\end{equation}
so the resulting expressions for the kinetic term and potential are,
 \begin{equation}\label{omega22}
 \omega(\phi)=-\frac{2\dot{H}}{\kappa^2-\frac{\kappa ^2 \rho_c- \sqrt{-12 H^2 \kappa ^2 \rho_c+\kappa ^4 \rho_c^2}}{ \rho_c}}-(1+w)\rho_0e^{-3(1+w_m)\int f(t)\mathrm{d}t},
\end{equation}
\begin{equation}\label{potential122}
V(\phi)= \frac{3H^2}{\kappa^2-\frac{\kappa ^2 \rho_c- \sqrt{-12 H^2 \kappa ^2 \rho_c+\kappa ^4 \rho_c^2}}{2 \rho_c}}+\frac{\dot{H}}{\kappa^2-\frac{\kappa ^2 \rho_c- \sqrt{-12 H^2 \kappa ^2 \rho_c+\kappa ^4 \rho_c^2}}{ \rho_c}}-\frac{(1-w)}{2}\rho_0e^{-3(1+w_m)\int f(t)\mathrm{d}t}\, ,
\end{equation}
where we used $P_m=w_m\rho_m$. Basically, the equations (\ref{omega22}) and (\ref{potential122}) generalize the reconstruction method of Refs. \cite{scalarrecon0,scalarrecon1,scalarrecon2}, in the LQC framework. Indeed, the classical limit can be obtained when $\rho_c\to \infty$, in which case, as it can be easily checked, the equations (\ref{omega22}) and (\ref{potential122}) become,
\begin{equation}\label{omega22classlim}
 \omega(\phi)=-\frac{2\dot{H}}{\kappa^2}-(1+w)\rho_0e^{-3(1+w_m)\int f(t)\mathrm{d}t},
\end{equation}
\begin{equation}\label{potential122classlim}
V(\phi)= \frac{3H^2}{\kappa^2}+\frac{\dot{H}}{\kappa^2}-\frac{(1-w)}{2}\rho_0e^{-3(1+w_m)\int f(t)\mathrm{d}t}\, ,
\end{equation}
which are identical to the classical expressions in Refs. \cite{scalarrecon0,scalarrecon1,scalarrecon2}.

Let us demonstrate how the reconstruction method we propose works using a simple example, which we take from \cite{scalarrecon0} in order to compare directly, since the limit $\rho_c\to \infty$ must yield the same results. Consider the case,
\begin{equation}\label{example1}
f(\phi )=H_0+\frac{H_1}{\phi^n}\,
\end{equation}
so by using Eqs. (\ref{omega22}) and (\ref{potential122}), the kinetic term reads,
\begin{equation}\label{kineticresult}
\omega(\phi)=-e^{-3 (1+w_m) \left(H_0 \phi +\frac{H_1 \phi ^{1-n}}{1-n}\right)} (1+w_m) \rho_m+\frac{2 H_1 n \phi ^{-1-n}}{ \kappa ^2 -\frac{\kappa ^2 \rho_c-\sqrt{\kappa ^4 \rho_c^2-12 \kappa ^2 \rho_c \left(H_0+H_1 \phi ^{-n}\right)^2}}{ \rho_c}}
\end{equation}
and as $\rho_c\to \infty$, this becomes,
\begin{equation}\label{inftyomega}
\omega(\phi)\simeq \frac{2 H_1 n t^{-1-n}}{\kappa ^2}-e^{-3 \left(H_0 t+\frac{H_1 t^{1-n}}{1-n}\right) (1+w_m)} (1+w_m) \rho_m\, ,
\end{equation}
which is identical to the example 3 of Ref. \cite{scalarrecon0}. Also, the potential in the LQC case is,
\begin{align}\label{lqcpotential}
& V(\phi)=\frac{6 t^{-2 n} \left(H_1+H_0 t^n\right)^2 \rho_c}{\kappa ^2 \rho_c+\sqrt{\kappa ^2 \rho_c \left(-12 \left(H_0+H_1 t^{-n}\right)^2+\kappa ^2 \rho_c\right)}}\\ \notag &
-\frac{ H_1 n t^{-1-n}}{\kappa ^2-\frac{6 \kappa ^2\,t^{-2 n} \left(H_1+H_0 t^n\right)^2}{\kappa ^2 \rho_c+\sqrt{\kappa ^2 \rho_c \left(-12 \left(H_0+H_1 t^{-n}\right)^2+\kappa ^2 \rho_c\right)}}}+e^{-3 t \left(H_0+\frac{H_1 t^{-n}}{1-n}\right) (1+w_m)} \frac{(-1+w_m)}{2} \rho_m
\end{align}
and in the classical limit $\rho_c\to \infty$, this becomes,
\begin{equation}\label{classicallimit}
V(\phi)\simeq -\frac{ H_1 n t^{-1-n}}{\kappa ^2}+\frac{3 \left(H_0+H_1 t^{-n}\right)^2}{\kappa ^2}-e^{-3 \left(H_0 t+\frac{H_1 t^{1-n}}{1-n}\right) (1+w_m)} \frac{(1-w_m)}{2} \rho_m
\end{equation}
which is identical to the one obtained in Ref. \cite{scalarrecon0}. Now an important issue that needs to be addressed is the stability of the solution (\ref{solutionsoftheform}) towards perturbations. We formally address this issue in the following section.

Up to now we discussed the classical regime, where the root (\ref{rhodefcorrect}) was used, but let us see what occurs if we use the root (\ref{rhodefcorrectrev}) which describes the quantum regime. It is conceivable that the limit $\rho_c\to \infty$ will not yield the classical limit since this root describes the era for which the quantum effects are strong. Consider the case of Eq. (\ref{example1}), so by using the root (\ref{rhodefcorrectrev}), the corresponding scalar kinetic term reads,
\begin{equation}\label{kineticresultrev1}
\omega(\phi)=-e^{-3 (1+w_m) \left(H_0 \phi +\frac{H_1 \phi ^{1-n}}{1-n}\right)} (1+w_m) \rho_m+\frac{2 H_1 n \phi ^{-1-n}}{ \kappa ^2 -\frac{\kappa ^2 \rho_c+\sqrt{\kappa ^4 \rho_c^2-12 \kappa ^2 \rho_c \left(H_0+H_1 \phi ^{-n}\right)^2}}{ \rho_c}}\, .
\end{equation}
We can readily see how the quantum root affects the dynamics, since in the limit $\rho_c\to \infty$, which is the classical limit, the kinetic term becomes,
\begin{equation}\label{kineticresultrev2}
\omega(\phi)=-e^{-3 (1+w_m) \left(H_0 \phi +\frac{H_1 \phi ^{1-n}}{1-n}\right)} (1+w_m) \rho_m+\frac{2 H_1 n \phi ^{-1-n}}{ \kappa ^2 -\frac{\kappa ^2 \rho_c+\kappa ^2 \rho_c}{ \rho_c}}\, .
\end{equation}
which is clearly different from the classical result (\ref{inftyomega}). Also, the potential for the root (\ref{rhodefcorrectrev}) is,
\begin{align}\label{lqcpotentialrev}
& V(\phi)=\frac{6 t^{-2 n} \left(H_1+H_0 t^n\right)^2 \rho_c}{\kappa ^2 \rho_c-\sqrt{\kappa ^2 \rho_c \left(-12 \left(H_0+H_1 t^{-n}\right)^2+\kappa ^2 \rho_c\right)}}\\ \notag &
-\frac{ H_1 n t^{-1-n}}{\kappa ^2-\frac{6 \kappa ^2\,t^{-2 n} \left(H_1+H_0 t^n\right)^2}{-\kappa ^2 \rho_c+\sqrt{\kappa ^2 \rho_c \left(-12 \left(H_0+H_1 t^{-n}\right)^2+\kappa ^2 \rho_c\right)}}}+e^{-3 t \left(H_0+\frac{H_1 t^{-n}}{1-n}\right) (1+w_m)} \frac{(-1+w_m)}{2} \rho_m
\end{align}
and in the classical limit $\rho_c\to \infty$, this becomes,
\begin{align}\label{lqcpotential}
& V(\phi)\simeq \frac{6 t^{-2 n} \left(H_1+H_0 t^n\right)^2 \rho_c}{\kappa ^2 \rho_c-\kappa ^2 \rho_c}\\ \notag &
-\frac{ H_1 n t^{-1-n}}{\kappa ^2-\frac{6 \kappa ^2\,t^{-2 n} \left(H_1+H_0 t^n\right)^2}{-\kappa ^2 \rho_c+\kappa ^2 \rho_c }}+e^{-3 t \left(H_0+\frac{H_1 t^{-n}}{1-n}\right) (1+w_m)} \frac{(-1+w_m)}{2} \rho_m\, ,
\end{align}
which is divergent, so the classical limit cannot be accessed, as it was expected since the root (\ref{rhodefcorrectrev}) captures the bouncing point era, which is the quantum era.

\section{Dynamical Stability of the non-Canonical Scalar Solution at First Order}

An important issue that needs to be properly addressed is the stability of the solution (\ref{solutionsoftheform}), since it is conceivable that this solution is not the only solution of the resulting cosmological equations. Notice that we will not address the stability issue for the root (\ref{rhodefcorrectrev}), which describes the bouncing point, since we are interested in the stability of the solutions which have the correct classical limit, in order to compare the results with the classical stability analysis. We shall be interested in the linear stability of the solution (\ref{solutionsoftheform}) towards linear perturbations of the dynamical system that corresponds the cosmological equations. Apart from the FRW equations (\ref{eqnm}) and (\ref{classicaleqns}), there is an additional equation of motion obeyed by the scalar field,
\begin{equation}\label{firledobey}
\omega (\phi)\ddot{\phi}+\frac{1}{2}\omega '(\phi)\dot{\phi}^2+3H\omega (\phi)\dot{\phi}+V'(\phi)=0\, ,
\end{equation}
where the prime denotes differentiation with respect to $\phi$. We introduce the following variables,
\begin{equation}\label{dynsvars}
X=\dot{\phi},\,\,\,Y=\frac{f(\phi )}{H}\, ,
\end{equation}
so the FRW equations (\ref{eqnm}), (\ref{classicaleqns}) and (\ref{dynsvars}) can be written as follows,
\begin{align}\label{perteqns}
& \frac{\mathrm{d}X}{\mathrm{d}N}=-3 (X-Y)+\frac{6 Y f'(t)}{\kappa ^2 \rho_c}+\frac{Y \left(9 f(t)^2-6 X^2 f'(t)\right)}{\kappa ^2 \rho_c}+\frac{f''(t)}{H(t) f'(t)}-\frac{X^2 f''(t)}{2 H(t) f'(t)}\\ \notag &
\frac{\mathrm{d}Y}{\mathrm{d}N}=\frac{X f'(t)}{H(t)^2}-\frac{X^2 Y f'(t)}{H(t)^2}\, ,
\end{align} 
where we used the $e-$foldings number $N=\ln a$, and also $\frac{\mathrm{d}}{\mathrm{d}N}=H^{-1}\frac{\mathrm{d}}{\mathrm{d}t}$. By taking the limit $\rho_c\to \infty$, the dynamical system (\ref{perteqns}) becomes,
\begin{align}\label{perteqnsperitto}
& \frac{\mathrm{d}X}{\mathrm{d}N}=-3 (X-Y)+\frac{f''(t)}{H(t) f'(t)}-\frac{X^2 f''(t)}{2 H(t) f'(t)}\\ \notag &
\frac{\mathrm{d}Y}{\mathrm{d}N}=\frac{X f'(t)}{H(t)^2}-\frac{X^2 Y f'(t)}{H(t)^2}\, ,
\end{align}
and as it can be seen, the resulting dynamical system (\ref{perteqnsperitto}) is identical to the classical dynamical system of \cite{scalarrecon3}, in the limit $\rho_c\to \infty$, and this validates that our analysis provides the correct expressions in the LQC case. 

Proceeding to find the stability conditions, the solution (\ref{solutionsoftheform}) corresponds to the values $(X,Y)=(1,1)$, so in order to study the stability of the ``fixed point'' $(X,Y)=(1,1)$, we consider the following linear perturbations,
\begin{equation}\label{pertactual}
X=1+\delta X,\,\,\,Y=1+\delta Y\, .
\end{equation}
By substituting the perturbations (\ref{pertactual}) in the dynamical system (\ref{perteqns}) and also by keeping first order terms of the variations $\delta X$ and $\delta Y$, we obtain the following dynamical system, 
\begin{equation}\label{perteqn2}
\left(
\begin{array}{c}
 \frac{\mathrm{d} X}{\mathrm{d}N} \\
 \frac{\mathrm{d} Y}{\mathrm{d} N} \\
\end{array}
\right)=\left(
\begin{array}{cc}
 -3-\frac{ H''(t)}{H(t) H'(t)}-\frac{12 H'(t)}{\kappa ^2 \rho_c} & 3+\frac{9 H(t)^2}{\kappa ^2 \rho_c} \\
 -\frac{H'(t)}{H(t)^2} & -\frac{H'(t)}{H(t)^2} \\
\end{array}
\right)\left(
\begin{array}{c}
 \delta X \\
 \delta Y \\
\end{array}
\right)\, .
\end{equation}
The dynamical system (\ref{perteqn2}) determines the evolution of the perturbations of the solution $(X,Y)=(1,1)$, and therefore determines the stability of the solution (\ref{sol1}). A direct comparison of the resulting perturbation equations (\ref{perteqn2}) to the ones obtained in Ref. \cite{scalarrecon3}, can show that in the limit $\rho_c\to \infty$, the LQC perturbation equations agree with the classical result of Ref. \cite{scalarrecon3}, in which case the dynamical system would read,
\begin{equation}\label{perteqn2peritto}
\left(
\begin{array}{c}
 \frac{\mathrm{d} X}{\mathrm{d}N} \\
 \frac{\mathrm{d} Y}{\mathrm{d} N} \\
\end{array}
\right)=\left(
\begin{array}{cc}
 -3-\frac{ H''(t)}{H(t) H'(t)} & 3 \\
 -\frac{H'(t)}{H(t)^2} & -\frac{H'(t)}{H(t)^2} \\
\end{array}
\right)\left(
\begin{array}{c}
 \delta X \\
 \delta Y \\
\end{array}
\right)\, .
\end{equation}
The stability of the system is determined by the eigenvalues $m_{1,2}$ of the matrix $M$ appearing in Eq. (\ref{perteqn2}), which is,
\begin{equation}\label{m}
M=\left(
\begin{array}{cc}
 -3-\frac{ H''(t)}{H(t) H'(t)}-\frac{12 H'(t)}{\kappa ^2 \rho_c} & 3+\frac{9 H(t)^2}{\kappa ^2 \rho_c} \\
 -\frac{H'(t)}{H(t)^2} & -\frac{H'(t)}{H(t)^2} \\
\end{array}
\right)\, ,
\end{equation} 
and if these are negative, then the solution (\ref{solutionsoftheform}) is stable towards linear perturbations. If one of the eigenvalues of $M$ is positive, then the solution (\ref{solutionsoftheform}) is unstable towards linear perturbations, and if one of these is zero, we cannot conclude if the solution (\ref{solutionsoftheform}) is stable. In the Appendix we present the detailed functional form of the eigenvalues of the matrix $M$. In the following section we shall study several illustrative examples and we shall make use of the findings of this section, in order to compare the classical and the loop quantum cosmology pictures, for a given cosmological evolution.

\section{Realization of Various Cosmologies Using LQC Scalar Fields}

In this section we present some illustrative examples in order to demonstrate how the realization of various cosmological scenarios can be done by using LQC non-canonical scalar fields. Our presentation has another aim, namely to compare the stability of the solution (\ref{solutionsoftheform}) in the context of LQC, with the solution of the corresponding classical theory. Also, the transition from non-phantom evolution to phantom is possible. Finally, we also briefly discuss a cosmological evolution which classically leads to a Big Rip singularity. As we show, in the LQC case the Big Rip singularity cannot be accessed, a result which agrees with other approaches using LQC \cite{nobigrip1,nobigrip1a,nobigrip1b,nobigrip1c,nobigrip2,nobigrip3,nobigrip4}. For all the examples we shall present, we assume that no matter fluids are present apart from the scalar field, so $\rho_m=p_m=0$.

\subsection{Non-Phantom Evolution}

\subsubsection{Evolution Away from the Quantum Regime}

Consider the following cosmological evolution,
\begin{equation}\label{ex1}
H(t)=H_0+\frac{H_1}{t^{n}}\, ,
\end{equation}
with $n>0$, and $H_0,H_1>0$. In the previous section we discussed this result in order to see that the resulting expressions in the LQC and classical case coincide in the limit $\rho_c\to \infty$, but now we shall study this example in a more formal way and in detail. Effectively, since the classical limit should be retained, this analysis corresponds to the root (\ref{rhodefcorrect}). In the context of LQC, in order to avoid inconsistencies in the definition of the energy density, pressure and of the kinetic term, we must require that $H_0$ and $H_1$ are chosen in such a way so that for the following inequality is satisfied all times,
\begin{equation}\label{ex2}
\kappa ^2 \rho_c> 12 \left(H_0+H_1 t^{-n}\right)^2\, .
\end{equation}
This constraint can be violated only at very early times, so if $H_0,H_1\ll 1$, and also if these parameters are appropriately chosen, then inconsistencies can be avoided. When $t>1$ and at late times, the inequality (\ref{ex2}) always holds true, if $H_0,H_1\ll 1$. By using Eq. (\ref{omega22}), and also by making use of the solution (\ref{solutionsoftheform}), the kinetic term $\omega (t)$ reads,
\begin{equation}\label{omegat}
\omega (t)=\frac{2 H_1 n t^{-1-n} \rho_c}{\sqrt{\kappa ^2 \rho_c \left(-12 \left(H_0+H_1 t^{-n}\right)^2+\kappa ^2 \rho_c\right)}}\, ,
\end{equation} 
which is always positive, so the scalar field is non-phantom. Also the potential that realizes the cosmology (\ref{ex1}) reads,
\begin{equation}\label{potex1}
V(t)= -\frac{t^{-2 n} \rho_c \,H_1 n t^{-1+n}}{\sqrt{\kappa ^2 \rho_c \left(-12 \left(H_0+H_1 t^{-n}\right)^2+\kappa ^2 \rho_c\right)}}+\frac{t^{-2 n} \rho_c\, 6 \left(H_1+H_0 t^n\right)^2}{\kappa ^2 \rho_c+\sqrt{\kappa ^2 \rho_c \left(-12 \left(H_0+H_1 t^{-n}\right)^2+\kappa ^2 \rho_c\right)}}\, .
\end{equation}
The EoS appearing in Eq. (\ref{eosdefcorrect}) can easily be computed and it reads,
\begin{equation}\label{bott}
w_{eff}=-1+\frac{H_1 n t^{-1-n}}{3 \left(H_0+H_1 t^{-n}\right)^2}+\frac{H_1 n t^{-1-n} \kappa ^2 \rho_c}{3 \left(H_0+H_1 t^{-n}\right)^2 \sqrt{\kappa ^2 \rho_c \left(-12 \left(H_0+H_1 t^{-n}\right)^2+\kappa ^2 \rho_c\right)}}\, ,
\end{equation}
and since $H_0,H_1\ll 1$, and also $n>1$, at early times, for $t\simeq 10^{-30}$sec, $\kappa^2\rho_c\simeq 10^{10}$sec$^{-1}$, and $H_0\sim H_1\sim 10^{-20}$sec$^{-1}$, the EoS is approximately equal to $-1+\epsilon $, with $0<\epsilon<1$, thus a quintessential acceleration epoch is described. Also, at late times, the EoS approaches the de Sitter value $-1$. The eigenvalues of the matrix $M$, $m_{1,2}$ can easily be calculated, and it can be easily shown that the stability properties of the solution (\ref{solutionsoftheform}) for the LQC case, are similar to the classical case.

\subsubsection{Evolution Near the Quantum Regime}

Having discussed the dynamics away from the bouncing point, we now discuss the analysis near the bouncing point which corresponds to the root (\ref{rhodefcorrectrev}). In this case, the kinetic term is,
\begin{equation}\label{omegatrev}
\omega (t)=-\frac{2 H_1 n t^{-1-n} \rho_c}{\sqrt{\kappa ^2 \rho_c \left(-12 \left(H_0+H_1 t^{-n}\right)^2+\kappa ^2 \rho_c\right)}}\, ,
\end{equation}
and as it is expected, the classical limit $\rho_c \to \infty $ does not yield the same result as in Eq. (\ref{omegat}). Correspondingly, the potential in this case is,
\begin{equation}\label{potex1rev}
V(t)= -\frac{t^{-2 n} \rho_c \,H_1 n t^{-1+n}}{\sqrt{\kappa ^2 \rho_c \left(-12 \left(H_0+H_1 t^{-n}\right)^2+\kappa ^2 \rho_c\right)}}-\frac{t^{-2 n} \rho_c\, 6 \left(H_1+H_0 t^n\right)^2}{-\kappa ^2 \rho_c+\sqrt{\kappa ^2 \rho_c \left(-12 \left(H_0+H_1 t^{-n}\right)^2+\kappa ^2 \rho_c\right)}}\, ,
\end{equation}
which again is different from the one given in Eq. (\ref{potex1}). Notice that at $t=0$ the Hubble rate becomes unbounded, and this is an important issue which we discuss later on when we study Big Rip singularities and unbounded Hubble rates.

\subsection{Oscillating Cosmologies and Non-phantom to Phantom Transitions}

\subsubsection{Evolution Away from the Quantum Regime}

Consider the following oscillating cosmological evolution,
\begin{equation}\label{ex11}
H(t)=H_0+H_i\sin \omega t\, ,
\end{equation}
with  $H_0,H_i>0$. As in the previous case, in order to have
\begin{equation}\label{ex21}
\kappa ^2 \rho_c> 12 \left(H_0+H_i \sin \omega t\right)^2\, ,
\end{equation}
for all cosmic times, $H_0$ has to be chosen in such a way so that $\kappa ^2\rho_c\gg H_0$. Then by combining Eqs. (\ref{omega22}) and (\ref{solutionsoftheform}), the kinetic term $\omega (t)$ reads in this case,
\begin{equation}\label{omegat1}
\omega (t)=\frac{2 H_i \rho_c \omega  \cos (t \omega )}{\sqrt{\kappa ^2 \rho_c \left(\kappa ^2 \rho_c-12 (H_0+H_i \sin(t \omega ))^2\right)}}
\, ,
\end{equation} 
so the appearance of the cosine in the above expression makes the kinetic term positive or negative in a periodic way and effectively, the non-canonical scalar field transits from phantom to non-phantom states. Moreover, the scalar potential that realizes the oscillating cosmology (\ref{ex11}) is,
\begin{equation}\label{potex11}
V(t)=\frac{H_i \rho_c \omega  \cos(t \omega )}{\sqrt{\kappa ^2 \rho_c \left(\kappa ^2 \rho_c-12 (H_0+H_i \sin (t \omega ))^2\right)}}+\frac{6 \rho_c(H_0+H_i \sin(t \omega ))^2}{\kappa ^2 \rho_c+\sqrt{\kappa ^2 \rho_c \left(\kappa ^2 \rho_c-12 (H_0+H_i \sin (t \omega ))^2\right)}}\, .
\end{equation}
Finally, the EoS of Eq. (\ref{eosdefcorrect}) for the cosmology (\ref{ex11}) reads,
\begin{equation}\label{bott1}
w_{eff}=-1-\frac{H_i \omega  \cos(t \omega )}{3 (H_0+H_i \sin(t \omega ))^2}-\frac{H_i \kappa ^2 \rho_c \omega  \cos(t \omega )}{3 (H_0+H_i \sin(t \omega ))^2 \sqrt{\kappa ^2 \rho_c \left(\kappa ^2 \rho_c-12 (H_0+H_i \sin(t \omega ))^2\right)}}\, ,
\end{equation}
from where it can be seen that the EoS transits from phantom to quintessential eras in a periodic way. In this case too by calculating the eigenvalues of the matrix $M$, it can be shown that the classical and LQC stability conditions have similar properties.

\subsubsection{Evolution Near the Quantum Regime}

Let us see the expressions of the kinetic term $\omega (t)$ and of the potential $V(t)$ corresponding to the root (\ref{rhodefcorrectrev}), which corresponds to the quantum regime. The kinetic term $\omega (t)$ reads in this case,
\begin{equation}\label{omegat1rev}
\omega (t)=-\frac{2 H_i \rho_c \omega  \cos (t \omega )}{\sqrt{\kappa ^2 \rho_c \left(\kappa ^2 \rho_c-12 (H_0+H_i \sin(t \omega ))^2\right)}}
\, ,
\end{equation} 
so the appearance of the cosine in the above expression makes the kinetic term positive or negative in a periodic way and effectively, the non-canonical scalar field transits from phantom to non-phantom states. Moreover, the scalar potential that realizes the oscillating cosmology (\ref{ex11}) is,
\begin{equation}\label{potex11rev}
V(t)=\frac{H_i \rho_c \omega  \cos(t \omega )}{\sqrt{\kappa ^2 \rho_c \left(\kappa ^2 \rho_c-12 (H_0+H_i \sin (t \omega ))^2\right)}}-\frac{6 \rho_c(H_0+H_i \sin(t \omega ))^2}{-\kappa ^2 \rho_c+\sqrt{\kappa ^2 \rho_c \left(\kappa ^2 \rho_c-12 (H_0+H_i \sin (t \omega ))^2\right)}}\, .
\end{equation}

\subsection{Big Rip Singularity and Unbounded Hubble Rates}

It is a well known fact in the context of LQC the Big Rip finite time singularities are avoided, see for example \cite{nobigrip1,nobigrip1a,nobigrip1b,nobigrip1c,nobigrip2,nobigrip3,nobigrip4,haronobigrip}, and therefore, in the case of scalar field LQC, the same result should be obtained. In this case we discuss this issue by using a characteristic example, and as we show, the Big Rip singularity cannot be obtained by LQC scalar field theory, at least when the root (\ref{rhodefcorrect}). This issue however requires careful considerations because near a Big Rip singularity the Hubble rate becomes unbounded and therefore the whole framework must be changed as we will see. This issue of unbounded Hubble rates was firstly addressed in Ref. \cite{rev1}, and there the Friedmann equation (\ref{holcor1}) was modified in order to describe theories with unbounded Hubble rates. Let us firstly demonstrate the problem in the LQC framework of Eq. (\ref{holcor1}) and then we discuss how should the problem addressed in a correct way.

Consider the following cosmological evolution,
\begin{equation}\label{nbr1}
H(t)=\frac{\frac{1}{2}\rho_ct}{\frac{3}{4}\rho_ct^2+1}+f_0\left(t-t_s \right)^{-\alpha}\, ,
\end{equation}
with $\alpha$ a real parameter, the value of which we specify now. Also $f_0>0$ and $t_s$ is a late time instance. Depending on the values of $\alpha$, classically the following types of finite time singularities can occur,
\begin{itemize}\label{lista}
\item For $\alpha<-1$, a Type IV singularity occurs at $t_s$.
\item For $-1<\alpha<0$, a Type II singularity occurs at $t_s$.
\item For $0<\alpha<1$, a Type III singularity occurs at $t_s$.
\item For $\alpha>1$, a Type I, or so called Big Rip singularity occurs at $t_s$, 
\end{itemize}
where we used the classification of finite time singularities given in Ref. \cite{Nojiri:2005sx}. Hence in the case that $\alpha>1$, the classical cosmological evolution would result to a Big Rip singularity. So from now on, we assume that $\alpha>1$ and furthermore in order to avoid complex values in the scale factor and in the Hubble rate, the parameter $\alpha$ is assumed to be of the form, $\alpha=2n/(2m+1)$, with $n,m>0$ integers. In the classical theory, the kinetic factor of the non-canonical scalar field theory that can realize the cosmology (\ref{nbr1}) is equal to \cite{scalarrecon0},
\begin{equation}\label{omegacl}
\omega (\phi)= \frac{2 f_0 \alpha  (-t_s+\phi )^{-1-\alpha }}{\kappa ^2}+\frac{3 \rho_c^2 \phi ^2}{2 \kappa ^2 \left(1+\frac{3 \rho_c \phi ^2}{4}\right)^2}-\frac{\rho_c}{\kappa ^2 \left(1+\frac{3 \rho_c \phi ^2}{4}\right)}\, ,
\end{equation}
and also the corresponding potential $V(\phi)$ is equal to \cite{scalarrecon0},
\begin{equation}\label{potcl}
V(\phi )=\frac{-f_0 \alpha  (-t_s+\phi )^{-1-\alpha }-\frac{3 \rho_c^2 \phi ^2}{4 \left(1+\frac{3 \rho_c \phi ^2}{4}\right)^2}+\frac{\rho_c}{2 \left(1+\frac{3 \rho_c \phi ^2}{4}\right)}+3 \left(f_0 (-t_s+\phi )^{-\alpha }+\frac{\rho_c \phi }{2 \left(1+\frac{3 \rho_c \phi ^2}{4}\right)}\right)^2}{\kappa ^2}\, .
\end{equation}
Finally, the classical EoS is equal to \cite{scalarrecon0},
\begin{equation}\label{eoscl}
w_{eff}=-1+\frac{2 (t-t_s)^{-1+\alpha } \left(2 (t-t_s)^{1+\alpha } \rho_c \left(-4+3 t^2 \rho_c\right)+f_0 \alpha  \left(4+3 t^2 \rho_c\right)^2\right)}{3 \left(2 t (t-t_s)^{\alpha } \rho_c+f_0 \left(4+3 t^2 \rho_c\right)\right)^2}\, .
\end{equation}
For the solution (\ref{solutionsoftheform}), since $\phi=t$, the classical kinetic term (\ref{omegacl}) and the potential (\ref{potcl}) strongly diverge at the Big Rip time instance $t=t_s$, and also the classical pressure and the classical energy density diverge at $t=t_s$. This means that in the classical description, the cosmological system reaches a Big Rip singularity. However, the LQC picture is different as we now evince. In this case, the kinetic term $\omega (\phi)$ for the cosmological evolution (\ref{nbr1}) reads,
\begin{equation}\label{omegaq}
\omega (\phi )=\frac{2 \rho_c (-t_s+\phi )^{-1-\alpha } \left(2 \rho_c (-t_s+\phi )^{1+\alpha } \left(-4+3 \rho_c \phi ^2\right)+f_0 \alpha  \left(4+3 \rho_c \phi ^2\right)^2\right)}{\left(4+3 \rho_c \phi ^2\right)^2 \sqrt{\kappa ^2 \rho_c \left(\kappa ^2 \rho_c-12 \left(f_0 (-t_s+\phi )^{-\alpha }+\frac{2 \rho_c \phi }{4+3 \rho_c \phi ^2}\right)^2\right)}}\, ,
\end{equation} 
and the corresponding potential reads,
\begin{align}\label{corpotq}
&V(\phi )=\frac{\rho_c (-t_s+\phi )^{-2 \alpha }}{\left(4+3 \rho_c \phi ^2\right)^2} \left(-\frac{(-t_s+\phi )^{-1+\alpha } \left(2 \rho_c (-t_s+\phi )^{1+\alpha } \left(-4+3 \rho_c \phi ^2\right)+f_0 \alpha  \left(4+3 \rho_c \phi ^2\right)^2\right)}{\sqrt{\kappa ^2 \rho_c \left(\kappa ^2 \rho_c-12 \left(f_0 (-t_s+\phi )^{-\alpha }+\frac{2 \rho_c \phi }{4+3 \rho_c \phi ^2}\right)^2\right)}}\right. \\ \notag &
\left.+\frac{6 \left(2 \rho_c \phi  (-t_s+\phi )^{\alpha }+f_0 \left(4+3 \rho_c \phi ^2\right)\right)^2}{\kappa ^2 \rho_c+\sqrt{\kappa ^2 \rho_c \left(\kappa ^2 \rho_c-12 \left(f_0 (-t_s+\phi )^{-\alpha }+\frac{2 \rho_c \phi }{4+3 \rho_c \phi ^2}\right)^2\right)}}\right)\, .
\end{align}
Finally, the corresponding EoS is equal to,
\begin{align}\label{eosq}
& w_{eff}=-1+\frac{(t-t_s)^{-1+\alpha } \left(2 (t-t_s)^{1+\alpha } \rho_c \left(-4+3 t^2 \rho_c\right)+f_0 \alpha  \left(4+3 t^2 \rho_c\right)^2\right)}{3\left(2 t (t-t_s)^{\alpha } \rho_c+f_0 \left(4+3 t^2 \rho_c\right)\right)^2}\\ \notag &
+\frac{(t-t_s)^{-1+\alpha } \kappa ^2 \rho_c \left(2 (t-t_s)^{1+\alpha } \rho_c \left(-4+3 t^2 \rho_c\right)+f_0 \alpha  \left(4+3 t^2 \rho_c\right)^2\right)}{3\left(2 t (t-t_s)^{\alpha } \rho_c+f_0 \left(4+3 t^2 \rho_c\right)\right)^2 \sqrt{\kappa ^2 \rho_c \left(\kappa ^2 \rho_c-12 \left(f_0 (t-t_s)^{-\alpha }+\frac{2 t \rho_c}{4+3 t^2 \rho_c}\right)^2\right)}}\, .
\end{align}
For the solution (\ref{solutionsoftheform}), the appearance of the square root in Eqs. (\ref{omegaq}), (\ref{corpotq}) and (\ref{eosq}), makes the situation very different from the classical case. This is owing to the fact that before the time instance $t=t_s$ is reached, the argument of the square root becomes negative, and therefore this results to complex values for the kinetic term, the potential and the EoS. Also it can easily be checked that the energy density and the pressure become complex before the Big Rip singularity is reached, since the pressure and energy density depend on the potential and the kinetic term, as it can be seen in Eq. (\ref{presenrg}). 

However, at this point we need to make the crucial observation that the Hubble rate (\ref{nbr1}) becomes unbounded at finite late time, and as was shown in Ref. \cite{rev1}, the effective Hamiltonian of the LQC system is not polymerized in this case, and it is equal to,
\begin{equation}\label{nonpolham}
\mathcal{H}=-\frac{3 V}{8\pi G \beta_1}\sinh^2 (\beta_2p/2)+H_m\, ,
\end{equation}
with $H_m=\rho a^3$. The corresponding Friedmann equation is not the one appearing in Eq. (\ref{holcor1}), but the following,
\begin{equation}\label{holcorrevision}
H^2=\frac{8\pi G}{3}\rho\left (1+\frac{\rho}{4\rho_c} \right)\, ,
\end{equation}
and therefore the corresponding energy density solutions are not the ones appearing in Eqs. (\ref{rhodefcorrect}) and (\ref{rhodefcorrectrev}), and the correct solutions are the following,
\begin{equation}\label{solutionsrev}
\rho=\frac{-2\kappa ^2 \rho_c \pm 2\kappa  \sqrt{\rho_c} \sqrt{3 H(t)^2+\kappa ^2 \rho_c}}{ \kappa ^2}\, .
\end{equation}
By looking at the solutions (\ref{solutionsrev}), it can be seen that even in the case the Hubble rate is unbounded, like for example in the Big Rip case, no complex values occur in the square root, since everything is positive. Therefore, the claim that LQC effects are not sufficient to describe the evolution near certain Big Rip singularities is not correct, since the LQC framework is consistent even in this case and also the Rip singularity is avoided, as was shown in \cite{nobigrip1,nobigrip1a,nobigrip1b,nobigrip1c,nobigrip2,nobigrip3,nobigrip4}.

\section{The Canonical Scalar Field Case}

The canonical scalar field case can easily be addressed by using the same line of research we used in the previous sections. In the canonical scalar field case, the energy density and the pressure of the scalar field are equal to,
\begin{equation}\label{cansc1}
\rho=\frac{1}{2}\dot{\varphi}+V(\varphi),\,\,\, P=\frac{1}{2}\dot{\varphi}-V(\varphi)\, ,
\end{equation}
so in this section we shall be interested in solutions of the form $f(t)=H(t)$, which is different from the form given in Eq. (\ref{solutionsoftheform}), since $\varphi\neq t$ in this case. The procedure is the same however, hence, the canonical scalar field $\varphi$ as a function of the cosmic time $t$, can be found by solving the following differential equation,
\begin{equation}\label{solvdiffeqn123}
\dot{\varphi}^2=\rho+P=-\frac{2\dot{H}}{\kappa^2(1-\frac{2\rho}{\rho_c})}\, ,
\end{equation}
where we made use of the LQC pressure of Eq. (\ref{pr}), and $\rho$ is given in Eq. (\ref{correctdef}). In addition, the potential $V(\varphi (t))$ as a function of the cosmic time can be found by using Eqs. (\ref{pr}) and (\ref{correctdef}), and it reads,
\begin{equation}\label{potcan1}
V(\varphi (t))=\frac{6H^2}{\kappa^2\left( 1-\frac{\kappa ^2 \rho_c- \sqrt{-12 H^2 \kappa ^2 \rho_c+\kappa ^4 \rho_c^2}}{2 \kappa ^2\rho_c}\right)}+\frac{2\dot{H}}{\kappa^2(1-\frac{2\rho}{\rho_c})}\, .
\end{equation}
Eventually, by solving Eq. (\ref{solvdiffeqn123}), and inverting the argument, the function $t=t (\varphi)$ is found, and by replacing the resulting expression in the potential $V(\varphi (t))$, the final form of the potential $V(\varphi )$ can be found. 
\begin{figure}[h] \centering
\includegraphics[width=14pc]{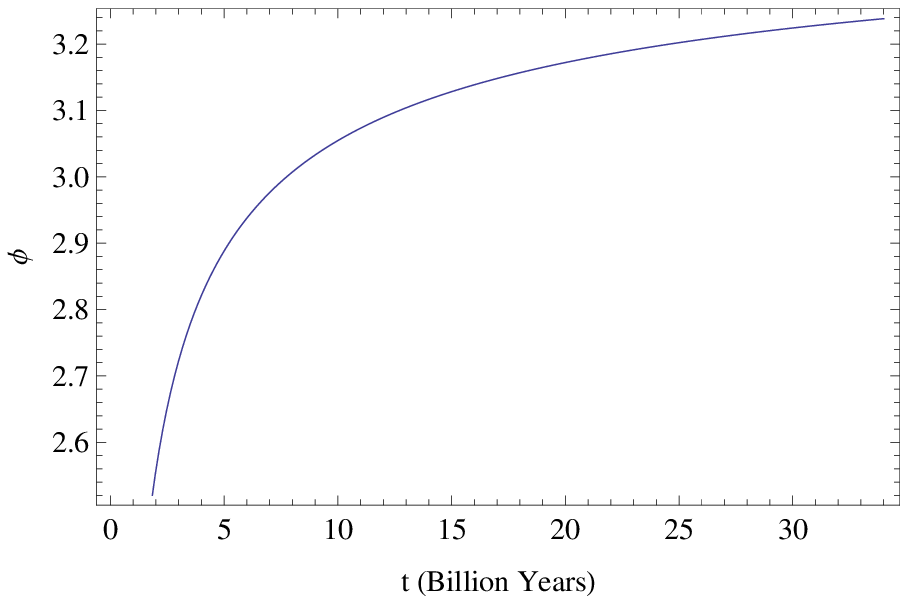}
\includegraphics[width=14pc]{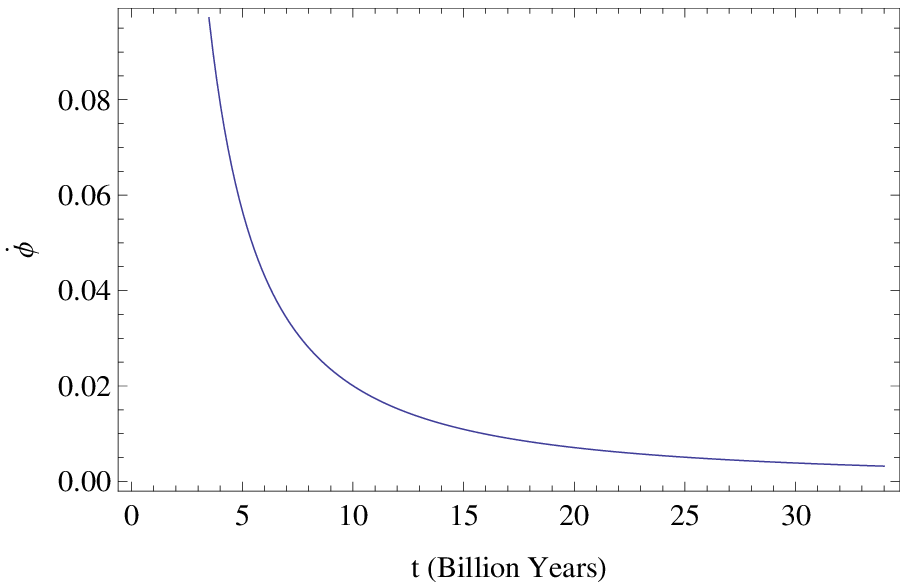}
\includegraphics[width=14pc]{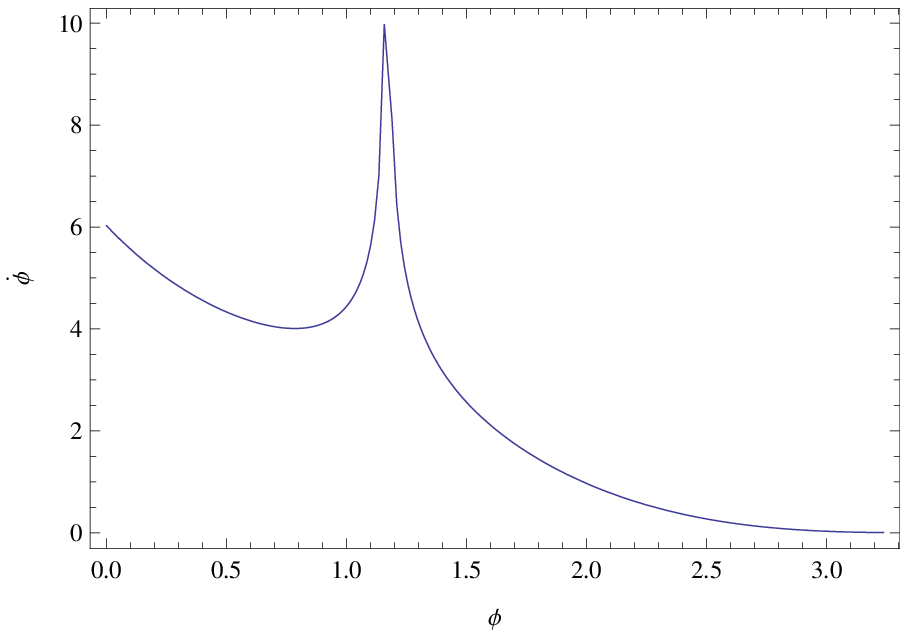}
\caption{The time dependence of the functions $\varphi (t)$ (left) and $\dot{\varphi} (t)$ (right) and also the phase plot $\dot{\varphi}-\varphi$ (bottom) for $\varphi (0.1)=0.001$ and $\kappa^2\rho_c=10$sec$^{-1}$, $H_0=0.1$sec$^{-1}$, $H_1=0.1$sec$^{-1+n}$}
\label{queen1}
\end{figure}
In order to demonstrate how this method works, consider the following cosmological evolution,
\begin{equation}\label{wscalefact}
a(t)=\left(\frac{3}{4} \kappa ^2(1+w)^2\rho_c t^2+1\right)^{\frac{1}{3 (1+w)}},\,\,\,H(t)=\frac{t (1+w) \kappa ^2 \rho_c}{2 \left(1+\frac{3}{4} t^2 (1+w)^2 \kappa ^2 \rho_c\right)}\, ,
\end{equation}
which corresponds to a perfect fluid with constant equation of state parameter $w$, so $P=w\rho$. By replacing the Hubble rate (\ref{wscalefact}) in the differential equation (\ref{solvdiffeqn123}), the canonical scalar $\varphi$ as a function of $t$ is equal,
\begin{equation}\label{ftcanon}
\varphi (t)=\frac{2 \,\, \mathrm{arcsinh}\left[\frac{1}{2} \sqrt{3} t (1+w) \kappa  \sqrt{\rho_c}\right]}{\sqrt{3} \kappa  \sqrt{(1+w) }}\, ,
\end{equation}
and also the potential $V(\varphi (t))$ reads,
\begin{equation}\label{varphivt}
V(\varphi (t))=-\frac{4 (-1+w) \rho_c}{4+3 t^2 (1+w)^2 \kappa ^2 \rho_c}\, .
\end{equation}
By inverting the function $\varphi (t)$ of Eq. (\ref{ftcanon}), we get,
\begin{equation}\label{frgr}
t=\frac{2 \sinh \left[\frac{\sqrt{3} \kappa  \sqrt{(1+w)} \varphi }{2}\right]}{\sqrt{3} (1+w) \kappa  \sqrt{\rho_c}}\, ,
\end{equation}
so by substituting in the potential of Eq. (\ref{varphivt}), the resulting potential $V(\varphi )$ reads,
\begin{equation}\label{resultingfinalpotential}
V(\varphi )=(1-w) \rho_c \,\mathrm{sech}\left[\frac{\sqrt{3} \kappa  \sqrt{(1+w)} \varphi }{2 }\right]^2\, .
\end{equation}
As it can be crosschecked, the resulting potential (\ref{resultingfinalpotential}) is identical to the one appearing in the literature, see for example \cite{polonos,ewing}. Hence, given the Hubble rate of an arbitrary cosmological evolution, by using Eqs. (\ref{correctdef}), (\ref{solvdiffeqn123}) and (\ref{potcan1}), we obtain the potential $V(\varphi)$, if the function $t=t (\varphi)$ can be found. 

\begin{figure} \centering
\includegraphics[width=14pc]{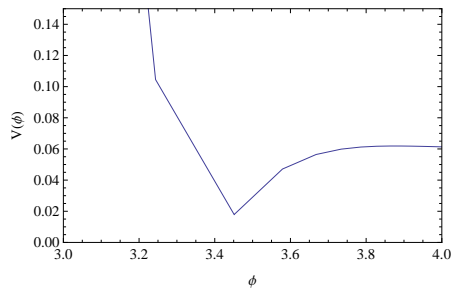}
\caption{The potential $V(\varphi)$ that generates the cosmology $H(t)=H_0+H_1/t^{n}$.}
\label{potentialf}
\end{figure}

It is worth discussing one of the examples we used in the case of a non-canonical scalar field, in order to have a concrete idea of the two cases. Consider for example the cosmological evolution of Eq. (\ref{ex1}), however it is not easy to obtain an analytic solution of the differential equation (\ref{solvdiffeqn123}), so we numerically solve it by using the initial condition $\varphi (0.1)=0.001$, and the values $\kappa^2\rho_c=10$sec$^{-1}$, $H_0=H_1=0.1$, so in Fig. \ref{queen1} we plot the time dependence of the functions $\varphi (t)$ (left) and $\dot{\varphi} (t)$ (right) and also the phase plot $\dot{\varphi}-\varphi$ (bottom). Also by using the data for $\varphi (t)$ and $V(\varphi (t))$, we can construct the parametric plot $V(\varphi)-\varphi$, which appears in Fig. \ref{potentialf}. 
Although it is not easy to have the analytic form of the potential $V(\varphi )$, we can fit the curve appearing in Fig. \ref{potentialf}, and the resulting approximate form of the potential is,
\begin{equation}\label{vfapprox}
V(\varphi )\sim c_1+c_2\varphi+c_3\varphi^2\, ,
\end{equation}
with $c_1\simeq 143.062$, $c_2=80.203$ and $c_3=11.122$. In order to see how accurate is the approximate expression for the potential (\ref{vfapprox}), we shall compute the EoS by using the expression given in Eq. (\ref{eosdefcorrect}), at various time instances, and we shall compare these values with ones obtained when the EoS is equal to,
\begin{equation}\label{approxexpress}
w_{eff}^{\varphi}=\frac{\frac{1}{2}\dot{\varphi}^2+V(\varphi (t))}{\frac{1}{2}\dot{\varphi}^2-V(\varphi (t))}\, ,
\end{equation} 
and by using the numerical results for $\dot{\varphi} (t)$ and $\varphi (t)$. Also the potential appearing in Eq. (\ref{approxexpress}), is the one we obtained in Eq. (\ref{vfapprox}). 
\begin{table*}[h]
\small
\caption{\label{comp}The values of the equation of state functions $w_{ap}$ and $w_{eff}$, for various cosmic times}
\begin{tabular}{@{}crrrrrrrrrrr@{}}
\tableline
\tableline
\tableline
Cosmic time (Billion Years) & $12.5$Gy & $13$Gy & $13.5$Gy & $14$Gy
\\\tableline
 $w_{eff}^{\varphi}$ & -0.999847 & -0.999856 &  -0.999863 &  -0.99987
\\\tableline
$w_{eff}$  &  -0.993239  & -0.993984 & -0.994623 &   -0.995175
\\
\tableline
\tableline
 \end{tabular}
\end{table*}
In Table \ref{comp}, we compare the EoS of Eq. (\ref{eosdefcorrect}) with the one appearing in Eq. (\ref{approxexpress}), and as it can be seen, the differences between the analytical expression and the numerical one, can be found at the second decimal point of the corresponding values, so the numerical approximation (\ref{vfapprox}) is relatively successful.

\section{Conclusions}

In this article we extended the scalar-tensor reconstruction techniques for realizing cosmological evolutions in the context of LQC. We presented the basic equations that constitute the LQC reconstruction method and we discussed the limitations of the method. Several examples were presented in order to demonstrate how the method works and also to show the new constraints that the LQC framework brings along. As we showed, it is possible to realize various cosmological scenarios and particularly certain features of a viable cosmology can be generated, for example the late and early-time acceleration era, phantom or quintessential evolution and even transitions between phantom and quintessential accelerations. As we showed the energy density has two branches of solutions with one yielding the classical limit and the other capturing the quantum phenomena. We discussed how the reconstruction method works in both these cases. We also addressed the case that the Hubble rate can be come unbounded at finite time. This issue is non-trivial and by adopting the method we used in the previous sections lead to inconsistencies. However, by using the right theoretical context presented in Ref. \cite{rev1}, the inconsistencies do not occur and formally the Rip singularities can be avoided.

In the case of non-canonical scalar fields, we also addressed the stability issue of the solution we proposed, and in all cases, the stability conditions are direct generalizations of the classical stability conditions, with the two coinciding in the classical limit $\rho_c\to \infty$. We also discussed the canonical scalar field case, and we studied the case of a perfect fluid with constant equation of state parameter $w$ and also we performed a numerical analysis for an example that was difficult to study analytically. 

A direct promising extension of the scalar-tensor LQC reconstruction method we proposed in this paper, is to use several scalar fields. This extension will provide a framework in which several cosmological scenarios could be realized, and also there is always the possibility for some scalar fields to be phantom and with the rest being non-phantom. Also the appearance of several scalars offers more freedom in realizing various evolution scenarios, so this theoretical extension should be worked out in detail in a future work.

\section*{Acknowledgments}

This work is supported by Min. of Education and Science of Russia (V.K.O).

\section*{Appendix: The Detailed Form of the Eigenvalues of $M$}

The detailed form of the eigenvalues of the matrix $M$ defined in Eq. (\ref{m}), is given below,
\begin{align}\label{m1}
& m_{1,2}=\frac{1}{2 \kappa ^2 \rho_c H(t)^3 H'(t)}\left(-3 \kappa ^2 \rho_c H(t)^3 H'(t)-\kappa ^2 \rho_c H(t) H'(t)^2-12 H(t)^3 H'(t)^2-\kappa ^2 \rho_c H(t)^2 H''(t)\pm \sqrt{Q(t)}\right)
\end{align}
where $Q(t)$ is,
\begin{align}\label{qt}
& Q(t)=\left(3 \kappa ^2 \rho_c H(t)^3 H'(t)+\kappa ^2 \rho_c H(t) H'(t)^2+12 H(t)^3 H'(t)^2+\kappa ^2 \rho_c H(t)^2 H''(t)\right)^2\\ \notag & -4 \kappa ^2 \rho_c H(t)^3 H'(t) \left(6 \kappa ^2 \rho_c H(t) H'(t)^2+9 H(t)^3 H'(t)^2+12 H(t) H'(t)^3+\kappa ^2 \rho_c H'(t) H''(t)\right)\, .
\end{align}

\end{document}